\documentclass{emulateapj}
\usepackage{amssymb,amsmath,bm}

\begin{document}

\title{Polarization of 21cm Radiation from the Epoch of Reionization}

\author{Daniel Babich \altaffilmark{1,2} and Abraham Loeb \altaffilmark{2,3}}

\altaffiltext{1}{Department of Physics, Harvard University, Cambridge, MA
02138; babich@physics.harvard.edu.}  

\altaffiltext{2}{Harvard-Smithsonian Center for Astrophysics, 60 Garden
Street, Cambridge, MA 02138} 

\altaffiltext{3}{Department of Astronomy, Harvard University, Cambridge, MA
02138; aloeb@cfa.harvard.edu.}

\begin{abstract}
We consider the polarization of 21cm line radiation from the epoch of
reionization due to both intrinsically polarized emission and secondary
mechanisms. We argue that Thomson scattering of the 21cm quadrupole by the
reionized universe is likely to produce the largest effect.  The 21cm
quadrupole is sourced by baryonic density fluctuations and the fluctuations
in the ionization fraction due to discrete \ion{H}{2} regions. Since
Thomson scattering produces only E-type polarization and is achromatic, its
unique fingerprint can in principle be separated from foregrounds
associated with polarized synchrotron emission which should not be
correlated with the cosmic signal. We estimate that Poisson fluctuations
of \ion{H}{2} regions at the end of reionization ($z_R \sim 6$--$20$)
produce a brightness temperature-polarization cross-correlation signal of
$\sim 0.1 - 0.3 \mbox{ mK}$ on angular scales of tens of arcminutes. This 
cross-correlation signal is within the instrument sensitivities of the future 
{\it Square Kilometer Array} (SKA) and close to the sensitivities of the forthcoming
{\it Mileura Widefield Array} (MWA) and {\it Low Frequency Array} (LOFAR).

\end{abstract}


\section{Introduction}

Measurements of the power-spectrum of 21cm brightness fluctuations can
potentially constrain the complex processes that reionized the universe
\citep{sunyaev75,hogan79,Scott,Tozzi,madau97,furlanetto04,Bar04c} as well as the basic cosmological
parameters \citep{loeb04,Bar04a,Bar05b}. By tuning low-frequency radio
arrays to different frequencies, one could observe the intergalactic medium
(IGM) at different redshift slices as it evolved through the epoch of
reionization. Observations of the temperature brightness power spectrum
will allow us to infer the baryonic power spectrum and the peculiar
velocity field at high redshifts, as well as the mean cosmic ionization
fraction and fluctuations in the ionization fraction due to discrete
\ion{H}{2} regions. Constraints on these quantities will test models of
reionization, in particular the spectra and evolution of the first sources
of ionizing radiation \citep{barkana01}. Several low-frequency arrays are
currently under construction for the purpose of mapping neutral hydrogen
during the epoch of reionization (http://space.mit.edu/eor-workshop/);
these include the {\it Primeval Structure Telescope}
(PAST)\footnote{http://astrophysics.phys.cmu.edu/~jbp}, the {\it Mileura
Widefield Array} (MWA)\footnote{http://web.haystack.mit.edu/arrays/MWA},
and the {\it Low Frequency Array} (LOFAR)\footnote{http://www.lofar.org},
that would establish the grounds for the construction of the {\it Square
Kilometer Array} (SKA)\footnote{http://www.skatelescope.org}.

Almost all theoretical work thus far has focused on the brightness
temperature fluctuations of the redshifted 21cm signal. The exact shape of
the brightness temperature fluctuation power spectrum during reionization
depends on the size and spatial distribution of the \ion{H}{2} regions and
should peak on the scale corresponding to the characteristic \ion{H}{2}
region size. On scales below this characteristic size the power in the
fluctuations should decline.  It is difficult to model this power spectrum
without introducing considerable astrophysical uncertainties, involving the
formation and spectra of the first stars and quasars \citep{Wyi03},
feedback effects that might impede the formation of future sources
\citep{Fur04}, and shadowing by compact low mass halos
\citep{Bar02,Sha04}. Ultimately, hydrodynamical simulations with radiative
transfer will be the most reliable method for following the evolution of
the \ion{H}{2} regions during the epoch of reionization. However, even
these will inevitably be sensitive to uncertain details about the processes
of star formation and quasar accretion which supply the ionizing photons.
At present the large dynamical range needed in order to follow reliably
radiative transfer in simulated boxes with periodic boundary conditions
\citep{Bar04d}, is still orders of magnitude beyond the values that
state-of-the-art codes \citep{Gnedin,Hern} are able to achieve.

In this paper we examine the polarization of the redshifted 21cm
radiation. The polarization could result from: {\it (i)}
intrinsic properties of the sources; and {\it (ii)} processes that polarize
the radiation as it travels towards the observer (secondary mechanisms). In
the first category we will analyze processes which produce a non-isotropic
population of the hyperfine triplet state, which upon spontaneous decay
could produce 21cm polarization. \citet{cooray04} already considered the
intrinsic polarization due to the Zeeman effect. They concluded that the
effect was most likely too small to be observed for realistic values of the
intergalactic magnetic fields at high redshifts. The second category of
polarization mechanisms is identical to the secondary processes which
produce polarization in the Cosmic Microwave Background (CMB)
\citep{zaldarriaga97b,hu00}. Here we will focus on the generation of
polarization out of the free-streaming 21cm anisotropies through Thomson
scattering by the reionized universe.

The polarization anisotropies carry important new information about
reionization.  For example, if the dominant source of polarization is
Thomson scattering in the reionized universe, then the 21cm polarization
would gauge the topology of ionized bubbles near the end of reionization as
well as the optical depth to electron scattering after the bubbles
overlaped. The latter measurement would help remove degeneracies among
cosmological parameters from the CMB data alone, and in particular
substantiate the inference about the optical depth from WMAP
\citep{spergel03}. The associated constraints on the growth and morphology
of \ion{H}{2} regions during the epoch of reionization
\citep{zahn05,mcquinn05}, can potentially remove the existing tension
between the conclusions drawn from the CMB data \citep{spergel03} and the
sizes of quasar \ion{H}{2} regions \citep{wyithe04a,Mes04}.

An additional reason to study the polarization of the 21cm signal is that
the expected low-frequency foreground is several orders of magnitude
brighter than the redshifted 21cm signal \citep{dimatteo02, oh03}. The
foregrounds are expected to be smooth in frequency space, while the signal
due to 21cm emission should fluctuate between narrowly spaced bands, and so
it was argued that the small signal can be extracted by taking the
differences between maps at narrowly separated frequency bands
\citep{zald04,Morales}. Since the foregrounds are expected to be polarized,
a definite prediction for the polarization, both the signal and
foreground, would further help us disentangle the signal from the
foregrounds. If the polarization signal is observable, it may be of a
particular parity type \citep[e.g. E-type or B-type, see][for a review of
polarization]{zaldarriaga97a}.  This additional information could also be
used to separate the signal from the foregrounds. Even if the polarization
signal is not expected to be observable by a particular experiment, we can
still use the prediction that any polarization detected by such an
experiment should not correlate with the inferred signal in order to reduce
the foreground contamination.

The organization of the paper is as follows. In \S 2 we review the possible
mechanisms which can produce a polarized signal. In \S 3 we present
numerical results for secondary polarization anisotropies due to Thomson
scattering in the reionized universe, which we expect to be the dominant
source of polarization. Finally, \S 4 summarizes our main conclusions.  We
adopt the standard $\Lambda$CDM cosmological model cosistent with the {\it
WMAP} \citep{spergel03} data ($\Omega_b = 0.044$, $\Omega_m = 0.27$,
$\Omega_v = 0.73$, $n = 1$, $\sigma_8 = 0.9$ and $h = 0.72$), and we show
results for different values of the reionization optical depth and the
redshift of reionization. For simplicity, we ignore helium and assume a
pure hydrogen plasma. We also focus on 21cm emission since the spin
temperature of hydrogen is likely to exceed the CMB temperature of $T_{\rm
CMB}=2.7\times (1+z)$K due to heating by X-rays; the heating is expected to
occur soon after the first sources light up and long before the universe
gets reionized, since it requires only $\la 10^{-2}$eV (instead of $\ga
10$eV for ionization) per baryon.

\section{Polarization Mechanisms}

As already mentioned, the mechanisms that are capable of producing
polarization in the 21cm line divide into two broad classes: {\it (i)}
intrinsic emission; and {\it (ii)} secondary mechanisms. In order to
produce polarized intrinsic emission, the triplet state of the hyperfine
splitting needs to be excited in a non-isotropic way (e.g. through pumping
by an anisotropic radiation field or splitting by a magnetic field which
introduces a preferred direction) and then the spontaneous decay of this
excited state will produce polarized light. It is also necessary that
collisional processes, which drive the system to thermal equilibrium, will
not be more rapid than the spontaneous decay and anisotropic triplet
pumping rates \citep{zygelman05}.

Even if the intrinsic emission is unpolarized, secondary mechanisms such as
Thomson or resonant scattering along that line-of-sight would produce
polarization in an analogous manner to the secondary production of
polarization in the CMB.  In this case, the polarization is sourced by an
incident quadrupole brightness moment in the scatterer's frame of
reference.  The optical depth to electron scattering inferred by WMAP $\tau
\sim 0.17 \pm 0.04$ \citep{spergel03}, implies that secondary scattering
could potentially produce a non-negligible effect.  The fundamental
difference between the scattered 21cm polarization and reionization bump
for the CMB polarization is the source of the incident quadrupole. The
incident quadrupole in the CMB case originates at the surface of last
scattering at the redshift of recombination ($z=10^3$) and is due to the
projection and free-streaming of the correlated CMB temperature
fluctuations. The free-streaming distance is always the horizon size as
seen by the scatterer. While the basic physical process is identical, these
details are changed in the 21 cm case. First, the free-streaming distance
from the emitter to the scatterer is not always the horizon size, but set
by the frequency of the observed line radiation.  The dominant sources of
the incident quadrupole during reionization are likely to be baryonic
density fluctuations in the \ion{H}{1} gas and ionization fraction
fluctuations based on the topology of \ion{H}{2} regions. The optical depth
for resonant 21cm absorption is typically smaller than a percent
\citep{Carilli,FurLoe} and will be ignored in our discussion.

\subsection{Intrinsic Emission}

Isotropic symmetry needs to be broken in order produce net polarization
(i.e. a preferred direction) in 21cm line emission from an ensemble of
hydrogen atoms.  There are two simple ways to break isotropy: {\it (i)}
Zeeman splitting of the triplet state due to a directional magnetic field;
and {\it (ii)} Wouthuysen-Field effect due to an anisotropic Ly$\alpha$
radiation field.  The possibility of the Zeeman effect has already been
considered by \cite{cooray04}. In order for the signal to be measurable by
SKA, a coherent field strength of $\ga 100\mu$G on Mpc scales is required
at $z\sim 6$--$10$ , with an energy density comparable to the CMB at that
time. Such a field strength is well above current Faraday-rotation limits
\citep{Kronberg} on the intergalactic field outside galaxy clusters
(keeping in mind the adiabatic enhancement factor of $(1+z)^2$ to high
redshifts), and is orders of magnitude above theoretical expectations from
astrophysical sources \citep{FurLoeb,Kulsrud}.

The Wouthuysen-Field effect is the mechanism which couples the gas
temperature to the hyperfine spin temperature through the absorption and
emission of Ly$\alpha$ photons \citep{wouthuysen52, field59}. There is a
finite probability for a $2p$ state, which was reached by Ly$\alpha$
absorption by the singlet $1s$ state, to spontaneously decay into the
triplet $1s$ state. This process will convert a fraction of the singlet
states to triplet states. When the Ly$\alpha$ scattering rate is high, the
spin temperature which determines the relative populations of the singlet
and triplet states, will equal the kinetic temperature of the gas (which
sets the ``color temperature'' of the Ly$\alpha$ photons). During
reionization the kinetic temperature of the gas is heated (e.g. by a small
level of X-rays produced by the first compact objects) to a value much
greater than the CMB temperature. The corresponding increase in the spin
temperature allows the 21cm line to be seen in emission during the late
stage of reionization.

If the Ly$\alpha$ radiation as seen by a hydrogen atom has a quadrupole
moment, then the 21cm line emitted by this atom will be polarized \citep[as
long as atomic collisions can be neglected; see][]{zygelman05}.  The basic
idea is that continuum radiation from the first high-redshift sources will
be redshifted into the blue wing of the Ly$\alpha$ line profile. This
radiation field could potentially be anisotropic because the ionizing
sources are highly clustered at high-redshift \citep{wyithe05}.
Unfortunately, the large optical depth near the Ly$\alpha$ resonance ($\sim
10^5$ for a neutral IGM) quickly isotropizes the incident radiation as it
is redshifted through the line profile by the cosmological expansion.
Therefore only the photons in the far blue wing of the Ly$\alpha$ line
profile are expected to produce a non-isotropic pumping of the triplet
state distribution. However, these photons are much less likely to scatter
than the isotropized photons near the center of the line.  Even in the
vicinity of a bright quasar, the anisotropy of the incident radiation would
induce polarization only for the neutral IGM within a thin skin outside the
quasar's \ion{H}{2} region whose spatial width is of order the
mean-free-path at the center of the thermally-broadenned Ly$\alpha$ line,
$\la 0.1 \mbox{ pc}$ \citep{wyithe04c}.  We conclude that pumping by an
anisotropic Ly$\alpha$ radiation field is unlikely to produce observable
intrinsically polarized 21cm line radiation.

\subsection{Secondary Mechanisms}

Thomson scattering by free electrons along the line-of-sight can polarize
intrinsically-unpolarized emission if the radiation field as seen by the
scattering electrons possesses a quadrupole moment. The inhomogeneous
distribution of the emitting \ion{H}{1} gas naturally generates a 21cm
quadrupole as the radiation free-streams towards the free electrons in the
reionized universe. The 21cm emission originates from the neutral fraction
of the IGM, and so fluctuations in the ionization fraction will also
produce fluctuations in the 21cm brightness.  There should be fluctuations
in the redshift of reionization along different directions due to large
scale inhomogeneities \citep{barkana04,wyithe04c}. Unfortunately, the
detailed correlation function and size distribution of \ion{H}{2} regions
depends strongly on uncertain details of the astrophysics of star and black
hole formation.  This dependence introduces inherent uncertainties into the
theoretical modelling of the ionization fraction power spectrum
\citep{furlanetto04,mcquinn05}.

The end of reionization, however, offers a special regime where robust
model-independent predictions can be made. This was demonstrated by Wyithe
\& Loeb (2004b) who derived the characteristic size of individual
\ion{H}{2} regions just before their final overlap based on the extended
Press-Schetcher theory. This derivation, which was based on the general
physical considerations of cosmic variance and light propagation delay,
eliminates much of the dependence on the astrophysical details.  We will
therefore incorporate this more robust approach and derive the 21cm
polarization power spectrum from it. For simplicity, we will assume that
Thomson scattering starts after the final overlap of the \ion{H}{2}
regions, since most of the Thomson scattering optical depth is naturally in
the regime where the filling fraction of ionized regions is substantial.
In our model, the 21cm quadrupole is sourced by the Poisson fluctuations of
the \ion{H}{2} regions at the end of reionization. This effect will produce
the largest signal, however for completeness in \S 3 we will also display
results from the baryon density fluctuation power spectrum.

The two Stokes' parameters $\bm{Q}$ and $\bm{U}$ of polarized radiation can
be decomposed into the rotationally invariant E-type and B-type fields
\citep{seljak96}. Due to parity invariance, perturbations which
rotationally transform as scalars and therefore lack any handedness, will
only produce E-type polarization. The 21cm quadrupole is sourced by baryon
density and ionization fraction fluctuations and should only produce E-type
polarization since the induced polarization is azimuthally symmetric about
the initial perturbation wavevector.  This is an important conclusion
because it can be used to eliminate polarized foreground emission and
instrumental systematics.

Propagation effects, such as gravitational lensing \citep{zaldarriaga98} and 
Faraday rotation \citep{kosowsky96,kosowsky05}, will convert the E-type 
polarization produced by Thomson scattering into B-type polarization. The 
amplitude of B-modes produced by gravitational lensing should not be 
significant \citep{zaldarriaga98}, 
however the effect of Faraday rotation may be quite substantial at the relevant 
low frequencies. It is difficult to estimate the properties of the appropriate 
rotation measures since they depend on the magnetic field amplitude and coherence 
structure, not just in the Galactic interstellar medium, but also in the high redshift IGM. 
Nevertheless the effect of Faraday rotation has a known frequency dependence 
and it is possible that the developed techniques for foreground removal may be
adapted to address this challenge. 

While calculating the effect of Thomson scattering on the incident 21cm
quadrupole we will ignore the inhomogeneites present in the distribution of
scattering electrons.  For simplicity, we will calculate the expected
polarization signal assuming a sudden uniform reionization. In reality,
prior to the end of reionization the ionization fraction is less than
unity, with full ionization only in localized \ion{H}{2} regions around
biased clusters of sources. Since the ionized bubbles contain free
electrons, they would also scatter the incident quadrupole and produce
polarization. The resulting visibility function will be inhomogeneous and
scattering will be able to produce B-mode polarization. In all calculations
of the effect for the CMB, the produced B-mode power spectrum was confined
to small scales and was several orders of magnitude smaller in amplitude
than the large scale E-mode polarization \citep{hu00}; we do not expect
this conclusion to differ for the 21cm line radiation, and so our working
assumption is that the large scale polarization only depends on the optical
depth of the homogeneous universe. 

In \S 3 we will demonstrate that the 
effect of assuming an instantaneous, instead of gradual, homogeneous 
reionization history is modest by showing that the polarization power spectrum
can be well approximated by a polarization power spectrum calculated from an 
instantaneous reionization history. Of course, the two power spectra are not 
identical; however, the appropriate power spectrum error bars, both cosmic 
variance and instrumental noise, must be included in making the comparison.
Similiarly it has been noted that the reionization history can be approximated
as instantaneous for low sensitivity CMB polarization measurements (such as
WMAP), but for upcoming nearly cosmic variance limited measurements 
(such as Planck) a better model of reionization is necessary \citep{Haiman03,
Holder03}. Since the introduced theoretical uncertainties should be modest compared
to the large instrumental noise and foreground contamination, we will use the 
assumption of instantaneous reionization in this paper.

As customary, we write down the polarization transfer function in the
line-of-sight formalism \citep{seljak96, hu97}, which is the Legendre
expansion of the E-type polarization induced by a single Fourier mode
perturbation,
\begin{equation}
   \Delta^E_{\it l}(k,\nu) = \frac{3}{4}\sqrt{\frac{(l+2)!}{(l-2)!}} \int^{\eta_R}_0 d\eta 
   \frac{g(\eta)}{\eta^2k^2} j_{\it l}(k\eta) \Pi(k,\eta,\nu),
\label{transfer}
\end{equation}
where $k$ is the primordial wavevector, ${\it l}$ is the observed Legendre
mode, and $\eta$ is conformal lookback time, defined as
\begin{equation}
\eta(z) = \int^z_0 \frac{dz'}{H(z')}.
\end{equation}
Here we also define the visibility function, which is the probability that
a photon last scattered at $\eta$, as
\begin{equation}
g(\eta) = \frac{d\tau}{d\eta}e^{-\tau(\eta)},
\end{equation} 
where
\begin{equation}
\frac{d\tau}{d\eta} = \sigma_T x_e(z) n_b (1+z)^2,
\end{equation}
$\sigma_T$ is the cross-section for Thomson scattering, $x_e(z)$ is the 
ionization fraction and $n_b$ is the average
baryon number density today.  The incident quadrupole on the scatterer is
$\Pi(k,\eta)$ and $\nu$ is the observed frequency of the 21cm line. Since
Thomson scattering is achromatic, the change in the observed frequency will
simply be due to the cosmological redshift, thus we can relate the observed
frequency to the rest frame frequency, $\nu_0 = 1.4 \mbox{ GHz}$, as $\nu =
\nu_0/(1+z_E)$, where $z_E$ is the emission redshift.

The smooth intergalactic \ion{H}{1} gas has a 21cm optical depth
\citep{field59,bharadwaj04,Bar04a},
\begin{equation}\label{HI_tau}
   \tau = \frac{3c^3 \hbar A_{10} x_{H} n_b (1+\delta) (1+z)^3}{16 k_B \nu^2_0 T_S H(z)}
   \left[1-\frac{(1+z)}{H(z)}\frac{dv_r}{dr}\right],
\end{equation}
where $A_{10}$ is the spontaneous emission coefficient of the hyperfine
transition, $T_S$ is the spin temperature that determines the relative
populations of the triplet and singlet states, $x_H$ is the neutral
fraction of the hydrogen gas, and $\delta$ is the baryonic overdensity.  
Here $dv_r/dr$ is the radial gradient of the line-of-sight
peculiar velocity of the \ion{H}{1} gas. The optical depth is proportional to the
path-length over which the a 21cm photon resonates with hydrogen atoms,
which is in turn inversely proprtional to the rate by which the Doppler
shift of the medium changes along its path.  Equation (\ref{HI_tau})
contains contributions to the optical depth from both the cosmological
expansion and the peculiar velocity of the gas.  The fluctuation in the
brightness temperature, calculated from the equation of radiative transfer,
is
\begin{eqnarray}
  \lefteqn{\delta T_b(z) = \frac{T_S - T_{\rm CMB}}{(1+z)} \tau,} \hspace{15mm} \\ 
 &\approx& 23 [1 + (1+\mu^2)\delta] x_H \hspace{25mm} \nonumber \\
  &\times&  \frac{(T_S - T_{\rm CMB})}{T_S} \left(\frac{1+z}{10}\right)^{1/2} \mbox{mK},
\end{eqnarray}
where $\mu$ is defined as $\mu = \hat{\bm{k}} \cdot \hat{\bm{n}}$, the cosine of the angle between
the wavevector of the primordial perturbation and the line of sight.
Since the secondary polarization mechanism is achromatic, we will simply
consider the dimensionless brightness temperature fluctuation, $\psi$,
which is defined as
\begin{equation}\label{temp_fluc}
  \psi = [1 + (1+\mu^2)\delta] x_H \frac{(T_S - T_{\rm CMB})}{T_S}.
\end{equation}
In the limit $T_S \gg T_{\rm CMB}$, which we expect to hold late in reionization, this can 
simply be expressed as $\psi = x_H [1+(1+\mu^2)\delta]$.

These correlated brightness temperature fluctuations will produce a
quadrupole incident upon the scatterers because free-streaming of
fluctuations transfers power from low Legendre modes to higher ones
\citep{hu97}. Performing this projection and free-streaming, the incident
dimensionless quadrupole can be expressed as \citep{zaldarriaga97b,hu00}
\begin{eqnarray}\label{quadrupole}
  \lefteqn{\Pi(k,\eta,\nu) = \int^{+1}_{-1} d\mu P_2(\mu) e^{i \mu k (\eta_E-\eta)} \psi(k,\eta_E),} 
   \hspace{20mm} \\
   &=& ~ j_2[k(\eta_E-\eta)] x_H(k,\eta_E) (1 + \delta(k,\eta_E)) \nonumber \\ 
   &-& ~ j_2''[k(\eta_E-\eta)] x_H(\eta_E) \delta(k,\eta_E), \hspace{15mm} 
\end{eqnarray}
where $\eta_E$ is simply the conformal distance to redshift of emission,
$z_E$. Each prime denotes a derivative of the spherical Bessel function;
the derivative term in equation (10) is produced by the peculiar velocity
term in equation (\ref{temp_fluc}).

In difference from the calculation of CMB anisotropies, we need to
define two separate transfer functions for the polarization produced by the
baryonic density fluctuations,
\begin{eqnarray}
   _{\delta}\Delta^E_{\it l}(k,\nu) &=& 
   \frac{3}{4}\sqrt{\frac{(l+2)!}{(l-2)!}} \int^{\eta_R}_0 d\eta 
   \frac{g(\eta)}{\eta^2k^2} j_{\it l}(k\eta) \hspace{10mm} \nonumber \\
   &\times& [~j_2[k(\eta_E-\eta)] - j_2''[k(\eta_E-\eta)]],  
\end{eqnarray}
and the ionization fraction fluctuations,
\begin{equation}
   _{x}\Delta^E_{\it l}(k,\nu) = \frac{3}{4}\sqrt{\frac{(l+2)!}{(l-2)!}} \int^{\eta_R}_0 d\eta 
   \frac{g(\eta)}{\eta^2k^2} j_{\it l}(k\eta) j_2[k(\eta_E-\eta)].  
\end{equation}
We must make this distinction since we cannot simply model the ionization
fraction fluctuations in terms of the baryonic density fluctuation or
primordial curvature perturbation.

The observed E-type polarization power spectrum can be expressed as 
\begin{eqnarray}
C^E_{\it l}(\nu) = \frac{2}{\pi} \int k^2 dk ~ [\bar{x}^2_H P_{\delta}(k)~ 
 [_{\delta}\Delta^E_{\it l}(k,\nu)]^2 \nonumber \\ 
+ P_{x}(k)~[_{x}\Delta^E_{\it l}(k,\nu)]^2],
\end{eqnarray}
where $P_{\delta}(k)$ is the baryonic power spectrum and $P_x(k)$ is the
ionization fraction power spectrum. Here we have defined $\bar{x}_H$ to 
be the average ionization fraction.
We will ignore the cross-correlation between $\delta$ and
$x_H$. The baryon density fluctuation can be expressed as
\begin{equation}
  \delta(\bm{k},\eta_E) = T(k) D(\eta_E) \zeta(\bm{k}),
\end{equation}
where $T(k)$ is the standard transfer function calculated by CMBFAST
\footnote{http://www.cmbfast.org/}, $D(\eta)$ is the linear theory growth function
and $\zeta(\bm{k})$ is the primordial curvature fluctuation produced by
inflation. Thus, the baryonic power spectrum can be expressed as
\begin{equation}
    P_{\delta}(k,\eta_E) = T^2(k) D^2(\eta_E) P_{\zeta}(k),
\end{equation}
where $P_{\zeta}(k)$ is the scale-invariant primordial curvature power
spectrum.

As previously mentioned, there will also be correlations in the ionization
fraction fluctuations.  The relative importance of the contributions from
the baryonic power spectrum and the ionization fraction power spectrum
depends on the mean ionization fraction as well as the topology of
\ion{H}{2} regions. The calculation of $P_x(k)$ is complicated by
uncertainties about the ionizing source population and the radiative
transfer of the ionizing radiation.  There have been recent attempts to
semi-analytically model this power spectrum based on the halo model
\citep{mcquinn05}, which divides the power spectrum into contributions from
the internal distribution of individual halos and the large scale
clustering of halos that are biased tracers of the linear baryonic matter
power spectrum \citep{cooray02}. The contribution from individual halos
introduces Poisson fluctuations in the \ion{H}{2} regions and dominates the
power spectrum on small scales. On large scales, the halos are a biased
tracer of the linear baryonic power spectrum. The transition between these
two regimes occurs near the characteristic bubble size.

\begin{figure}
\includegraphics[height = 10cm]{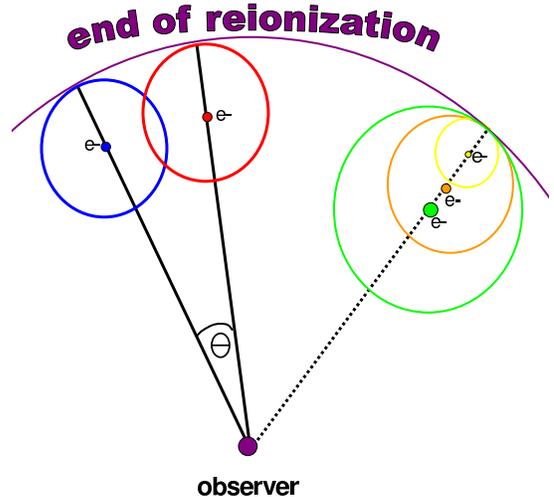}
\caption{A schematic illustration of the scattering geometry. Each electron
along the dotted (rightmost) line-of-sight scatters 21cm radiation that was
emitted just before the end of reionization at a redshift $z_R$ from a
spherical shell around it. All emitting shells are tangent to the bounding
shell around the observer. 21cm radiation emitted by the patchy \ion{H}{1}
distribution in these shells and scattered after reionization by the
electrons at the shell centers will show polarization at a wavelength of
$21{\rm cm}\times (1+z_R)$. The two solid lines-of-sight on the left which
are separated by an angle $\theta$ will show correlations in their
polarization amplitudes due to scattering electrons with intersecting
emission shells.  }
\label{diagram}	
\end{figure}

In this paper we focus on the contribution to the incident quadrupole from
Poisson fluctuations in the number of \ion{H}{2} regions. For simplicity
and definitiveness we will only include this effect at the end of
reionization, namely the surface of bubble overlap, where we can robustly
calculate the size and number density of \ion{H}{2} regions based on the
general cosiderations of cosmic variance and light propagation
delay\footnote{Note that inclusion of the light propagation delay is
essential for the calculation of the size of an \ion{H}{2} bubble at the
end of reionization, as viewed from the location of a scattering
electron. Any calculation of the typical bubble size in a spacelike
snapshot of the universe (as commonly done in the literature) is not
adequate for this purpose.}  \citep{wyithe04b}.  As long as the volume
filling fraction of ionized bubbles is small, the 21cm fluctuations are
small, and so most of the signal naturally originates around the time of
bubble overlap anyway. We will ignore the correlations of the ionization 
fraction due to large scale biasing of the \ion{H}{2} regions, as well as, 
cross correlations between the baryon density and the ionization fraction.
Biasing effects can only be important on scales above the characteristic
size of the \ion{H}{2} regions. However at the end of reionization the 
characteristic size of the \ion{H}{2} regions is extremely large 
($\sim 70$ comoving Mpc) \citep{wyithe04b}. On these scales the amplitude of the baryonic power 
spectrum is quite small so the biased \ion{H}{2} region power spectrum will 
be modest at the end 
of reionization. Besides these effects are primarily important on large 
angular scales, where the conversion of the temperature brightness 
fluctuations to the necessary quadrupole anisotropies is suppressed; 
therefore, we will find them to be below realistic instrument detection 
thresholds. 

The observed fluctuations in the ionization fraction are caused by
fluctuations in the local number of \ion{H}{2} regions. Poisson statistics
of point-like objects is described by the spatial two-point correlation
function,
\begin{equation}
  \langle \delta x(\bm{r}_1) \delta x(\bm{r}_2) \rangle =
\frac{1}{\bar{n}(\bm{r}_1)} \delta^{(3)}(\bm{r}_1-\bm{r}_2),
\end{equation}
where $\bar{n}$ is the average number density of \ion{H}{2} regions and $\delta x$ 
is the fluctuation in the ionization fraction. Since the
\ion{H}{2} regions have a finite size, we must convolve their density field
with a window function
\begin{equation}
   \delta x(\bm{r}_1) = \int d^3\bm{r}_2 W_R(|\bm{r}_1 - \bm{r}_2|) \delta x(\bm{r}_2),
\end{equation}
where we assume a Gaussian window function for computational simplicity
\begin{equation}
   W_R(r) = \frac{e^{-r^2/2R^2}}{\sqrt{(2\pi)^3} R^3}.
\end{equation}
The two-point correlation function is given by
\begin{eqnarray}
  \lefteqn{\langle \delta x(\bm{k}_1) \delta x^{*}(\bm{k}_2) \rangle = } \hspace{5mm} \nonumber \\
   && \int d^3 \bm{r}_1 d^3 \bm{r}_2 e^{i \bm{k}_1 \cdot \bm{r}_1} e^{-i \bm{k}_1 \cdot \bm{r}_1} 
      \int d^3 \bm{y}_1  d^3 \bm{y}_2 \nonumber \\ 
   &\times& \frac{e^{-|\bm{r}_1 - \bm{y}_1|^2/2R^2}}{\sqrt{(2\pi)^3} R^3}
      \frac{e^{-|\bm{r}_2 - \bm{y}_2|^2/2R^2}}{\sqrt{(2\pi)^3} R^3}
      \langle \delta x(\bm{y}_1) \delta x(\bm{y}_2) \rangle.
\end{eqnarray}
Simplifying we find
\begin{eqnarray}
\langle \delta x(\bm{k}_1) \delta x^{*}(\bm{k}_2) \rangle &=& \nonumber \\ 
       \frac{1}{\bar{n}} \int d^3 \bm{r}_1 
	& d^3 \bm{r}_2& e^{i \bm{k}_1 \cdot \bm{r}_1} e^{-i \bm{k}_1 \cdot \bm{r}_1} 
	\frac{e^{-D^2/R^2}}{\sqrt{(2\pi)^3}R^3},
\end{eqnarray}
where we have defined $\bm{D} = (\bm{r}_1 - \bm{r}_2)/2$. As $R \rightarrow 0$, the Gaussian window
function approaches $\delta^{(3)}(\bm{r}_1 - \bm{r}_2)$ and then we find $P_x(k)= 1/\bar{n}$, where 
we define the ionization fraction power spectrum, $P_x(k)$, through
$\langle \delta x(\bm{k}_1) \delta x^{*}(\bm{k}_2) \rangle = (2\pi)^3 \delta^{(3)}
(\bm{k}_1 - \bm{k}_2) P_x(k_1)$.
We can now analytically calculate the two-point function for finite bubble size
\begin{equation}
P_x(k) = \frac{1}{\bar{n}} e^{-k^2 R^2}.
\end{equation}

The translational invariance of $ \langle \delta x(\bm{r}_1) \delta x(\bm{r}_2)
\rangle $ guarantees that $ \langle \delta x(\bm{k}_1) \delta x^{*}(\bm{k}_2)
\rangle $ will be proportional to $\delta^{(3)}(\bm{k}_1 - \bm{k}_2)$. The
cutoff in the temperature brightness fluctuation power spectrum due to the
finite bubble size will cause a corresponding decay in the polarization
power spectrum for multipole index ${\it l} \ga \eta_R/R$.  Below this
scale, the afforementioned signal due to baryon fluctuations will come to 
dominate. However, as we will see in the next section, this contribution 
to the signal is far below the instrumental noise detection threshold.
Therefore we are justified in ignoring it.

We assumed $\bar{n}(\bm{r})$ to be constant because of the large-scale
homogeneity of the universe; the coordinate $\bm{r}$ is the comoving
distance on the hypersurface of simultaneity corresponding to the time of
reionization, not a coordinate labeling our past lightcone. This Poisson
fluctuation contribution to the polarization power spectrum corresponds to
the intersection (within the region defined by the bubble window function)
of the past light cones of the two electron scatterers, which both lie on
our past light cone, on the hypersurface of simultaneity corresponding to
the end of reionization. Figure \ref{diagram} schematically shows this
effect.  On the righthand side of the figure, scatterers along a given
line-of-sight see 21cm radiation from progressively larger shells because
the elapsed time between emission and scattering for an electron increases
for electrons closer to the observer. On the lefthand side of the figure,
the shells of 21 cm radiation emisson are displayed for two electrons along
two lines-of-sight separated by an angle $\theta$. These shells are
distinct from the observer's shell because they correspond to past light
cones of the electrons which intersect the hypersurface of simultaneity
when reionization ended at different physical positions. In the figure, the
shells for the two electrons (red and blue) intersect and therefore the
polarization produced along the two lines-of-sight will be correlated due
to Poisson fluctuations.

At the end of reionization, which we define to be the surface of bubble
overlap \citep{wyithe04b}, the mean distance between \ion{H}{2} regions is
equivalent to the characteristic size of the the bubbles. Thus $\bar{n}$,
the average bubble density, equals $1/V$, where $V = 4\pi R^3/3$ is the 
characteristic bubble volume. This is only true at the end of reionization; earlier on, during
the process of reionization, the characteristic bubble size is smaller than
the mean bubble separation. For our model, we find that the polarization
power spectrum is given by
\begin{equation}\label{HII_Cl}
C^E_{\it l}(\nu) = \frac{8 R^3}{3} \int k^2 dk e^{-R^2 k^2} ~ [_{x}\Delta^E_{\it l}(k,\nu)]^2. 
\end{equation}

Since the brightness temperature fluctuations are the source of the
quadrupole which produces polarization through Thomson scattering, the
polarization and temperature signals should be correlated. Actually only
the E-type polarization is correlated with the temperature fluctuations, as
parity prevents correlations with B-type polarization
\citep{zaldarriaga97a}. Depending on the characteristics of reionization,
this cross-correlation signal may be much easier to detect.  For this
reason the CMB temperature-polarization cross-correlation spectrum was
first published by DASI \citep{Kovac} and WMAP \citep{kogut03}. 
It is straightforward to adapt the above formalism. The temperature
transfer function is again divided into contributions from the baryon
density fluctuations
\begin{equation}\label{tran_t}
  _{\delta}\Delta^{T}_{\it l}(k,\nu) = j_{\it l}(k\eta_E) - j_{\it l}''(k\eta_E),
\end{equation}
and the ionization fraction fluctuations
\begin{equation}
  _{x}\Delta^{T}_{\it l}(k,\nu) = j_{\it l}(k\eta_E).
\end{equation}

Then the observed temperature polarization cross-correlation spectrum can be written as
\begin{eqnarray}\label{xl}
C^X_{\it l}(\nu) = \frac{2}{\pi} \int k^2 dk~  
[\bar{x}^2_H P_{\delta}(k) _{\delta}\Delta^E_{\it l}(k,\nu) _{\delta}\Delta^T_{\it l}(k,\nu) \nonumber \\
+ P_{x}(k) _{x}\Delta^E_{\it l}(k,\nu) _{x}\Delta^T_{\it l}(k,\nu)]. \hspace{10mm}
\end{eqnarray}

We have ignored effects due to the spectral response functions of the observing instruments. 
These are straightforward to incorporate into our formalism. Since we will find that the dominant
signal, which is produced by \ion{H}{2} region Poisson fluctuations, is rather featureless and
exponentially damped below scales of tens of arcminutes (${\it l} > 10^3$) reasonable instrument
bandwidths ($\Delta \nu \sim 0.4 \mbox{ MHz}$) will not substantially affect our results. 

\section{Results}

Based on the formalism developed in \S 2, we present numerical results for
a variety of emission and reionization redshifts in order to understand how
the signal depends on these parameters. We consider the simplest model in
which reionization is assumed to be sudden, so the ionization fraction is a
step function, and homogeneous, except for the effect of Poisson
fluctuations of \ion{H}{2} regions on the incident 21cm quadrupole. This
implies that the 21cm power spectrum will simply be due to correlations in
the baryonic density and that the gas is always completely neutral,
$\bar{x}_H = 1$, when the redshift of the emission is greater than the
redshift of reionization. When we observe emitted radiation from the
redshift of reionization, we include the contribution from the Poisson
fluctuations in the \ion{H}{2} region distribution. Once the universe
becomes reionized, it does so uniformly. Our approximate treatment ignores
scattering due to the patchiness of the universe while it is only partially
ionized. We adopt for our first treatment this simplified model because it
has the smallest number of free parameters; more complicated models with
uncertain astrophysical parameters can be considered in the future.

\begin{figure}
\plotone{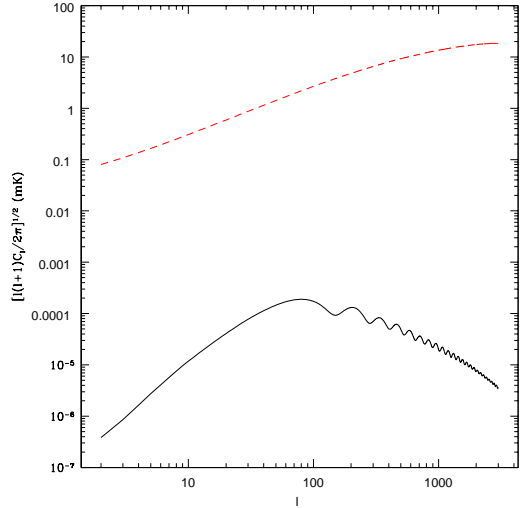}
\caption{The brightness temperature (red, dashed) and polarization (black,
solid) due to baryon density fluctuations for emission redshift of $z_E =
20$ and reionization redshift $z_R = 17$.}
\label{temp_elec}	
\end{figure}

Figure \ref{temp_elec} compares the temperature brightness power spectrum
($C^T_{\it l}$) and the polarization power spectrum ($C^E_{\it l}$) due to
baryon density fluctuations for $z_E = 20$, $z_R = 17$. The features
in $C^T_{\it l}$ and $C^E_{\it l}$ are quite different. In addition to
being at a lower amplitude, $C^E_{\it l}$ peaks and begins to oscillate and
decay at high ${\it l}$. The polarization transfer function, defined in
equation (\ref{transfer}), is the line-of-slight integral of two spherical
Bessel functions, weighted by the visibility function and additional
factors due to the polarization spin lowering operators. The properties of the spherical
Bessel function are well-known: when $x \ll 1$ $j_{\it l}(x) \sim x^l$, it
has a peak at $x \sim l$ and then at large $x$ it oscillates and decays as
$j_{\it l}(x) \sim \sin(x - \pi l/ 2)/x$. We will utilize these features to
understand the behaivor of $C^E_{\it l}$.

\begin{figure}
\plotone{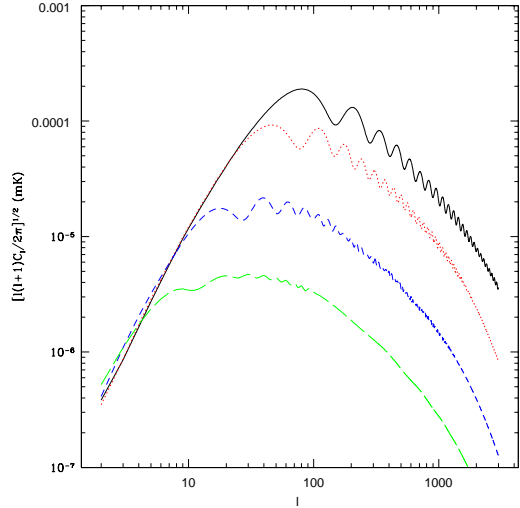}
\caption{The polarization power spectrum sourced by baryon density
fluctuations for an emission redshift $z_E = 20$ and reionization redshift
$z_R = 17$ (black, solid), $z_R = 15$ (red, dotted), $z_R = 10$ (blue,
dashed) and $z_R = 6$ (green, long dashed).}
\label{cl_hz}	
\end{figure}

Figure \ref{cl_hz} shows the polarization power spectrum for various
reionization redshifts, $z_R = 17, 15, 10, 6$ (top-to-bottom), and a single emission
redshift, $z_E = 20$. The basic physical processes become clear by
comparing these various power spectra. We will first consider the angular
position of the peak in the polarization power spectra.  The spherical
Bessel function $j_{\it l}(k \eta)$ peaks at ${\it l} \sim k \eta$ and
likewise $j_2[k(\eta_E-\eta)]$ at $2 \sim k(\eta_E -\eta)$. These two
constraints imply that the polarization power spectrum peaks at
\begin{equation}
{\it l}_{peak} \sim \frac{2 \eta_R}{\eta_E - \eta_R},
\label{peak}
\end{equation}
where $\eta_E$ and $\eta_R$ are the comoving distances to the redshifts of
emission and reionization, respectively. Here we have effectively assumed
an infinitely sharp visibility function. The inclusion of the altered
free-streaming term, due to peculiar velocities, only has a minor effect on
this argument.  Thus changing $z_R$, for a given $z_E$, will shift the
power spectrum features to higher ${\it l}$. This is clearly observed in
Fig. \ref{cl_hz} and numerically the peak locations agree with equation
(\ref{peak}).

The oscillatory features are due to the free-streaming of the monopole
brightness temperature fluctuations.  The incident quadrupole on a
scatterer, equation (\ref{quadrupole}), contains the spherical Bessel
function, $j_2[k(\eta_E-\eta)]$, which describes how a monopole fluctuation
of wavevector $k$ at $\eta_E$ becomes a quadrupole fluctuation at
$\eta$. The projection of this oscillatory function, the free-streaming
quadrupole, on the sky causes the features in Fig. (\ref{cl_hz}). These are
not the analogs of the acoustic peaks observed in the primary temperature
anistotropies of the CMB \citep{Bar05b}. Both are caused by projecting an
oscillatory function on the sky; however the acoustic oscillations
originate from pressure waves in the baryon-photon fluid prior to
recombination.

The decay of $C^E_{\it l}$ at high ${\it l}$ is explained by oscillatory
cancellation of the line-of-sight integration. The two spherical Bessel
functions would have had different phases and therefore integrated to zero
if the range of integration was infinite. When the wavelength of the
perturbation becomes comparable in size to the width of the visibility
function, this effect begins to become important. Since the visibility
function becomes progressively more peaked at higher reionization redshift,
progressively smaller intervals along the line-of-sight contribute to the
transfer function and therefore there is less damping. We observe this
phenomenon in Fig. \ref{cl_hz}, as the high ${\it l}$ tail decays less and
retains oscillatory features for the curves corresponding to the higher
reionization redshifts. If the change in the ionization fraction is not
sudden as we have assumed but rather gradual, then these oscillatory
features would be smoothed out.

\begin{figure}
\plotone{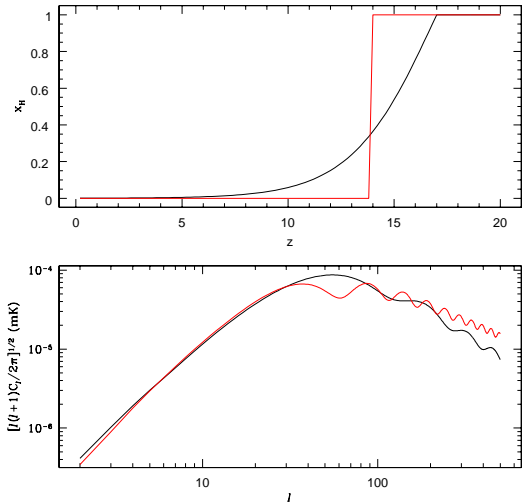}
\caption{Top Panel: The gradual and instantaneous reionization histories
appropriate for the polarization power spectra displayed in the botton panel. 
Bottom Panel: The polarization power spectra are displayed for the instantaneous 
and gradual reionization histories assuming that the 21cm fluctuations are 
sourced by density inhomogeneities. The color coding is the same in both panels.}
\label{diff}	
\end{figure}

In these examples we have assumed that reionization is instantaneous, while
in reality reionization must be gradual \citep{Fur04}. As mentioned in \S
2.2, the adopted reionization history will imprint a signature on the
polarization power spectrum. Next we demonstrate that the polarization
power spectrum sourced by baryon density fluctuations and produced for a
given gradual reionization history can be well approximated (within the
appropriate cosmic variance and instrumental noise errorbars) by an
instantaneous reionization history (with the total optical depth not being
necessarily the same in the two reionization histories).  In the bottom
panel of Fig. \ref{diff} the polarization power spectra sourced by baryon
density fluctuations are displayed for the instantaneous and gradual 
reionization histories plotted in the top panel. While there are differences 
between the two reionization histories, they are fairly modest compared to 
the much greater instrumental noise and model uncertainties. The
use of the simple model of instantaneous reionization is therefore well
justified for the scope of this paper.

\begin{figure}
\plotone{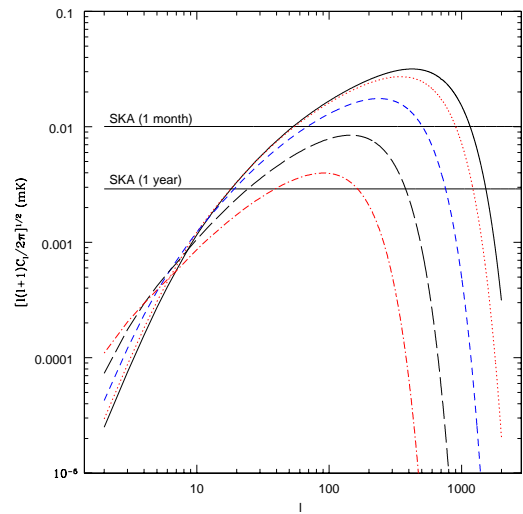}
\caption{The polarization power spectrum sourced by Poisson fluctuations of
\ion{H}{2} regions for an emission and reionization redshift $z_E = z_R = 30$ 
(black, solid), $z_E = z_R = 25$ (red, dotted), $z_E = z_R = 17$ (blue,
dashed), $z_E = z_R = 10$ (black, long dashed) and $z_E = z_R = 6$ (red,
dot dashed). The theoretical detection threshold of SKA is also shown for a 1
month integration time and a 1 year integration time with a bandwidth of 
$0.4 \mbox{ MHz}$.}
\label{poisson}	
\end{figure}

Figure \ref{poisson} displays the polarization power spectra sourced by
Poisson fluctuations of \ion{H}{2} region, for different redshifts of
emission and reionization: $z_E = z_R = 30$ 
(black, solid), $z_E = z_R = 25$ (red, dotted), $z_E = z_R = 17$ (blue,
dashed), $z_E = z_R = 10$ (black, long dashed) and $z_E = z_R = 6$ (red,
dot dashed). The theoretical detection threshold of SKA is also shown for a 1
month integration time and a 1 year integration time. We can understand the
basic features of the polarization power spectra by inspecting equation
(\ref{HII_Cl}). The amplitude of the polarization power spectra is
proportional to the characteristic volume of an \ion{H}{2} bubble at the
time of bubble overlap. If reionization completes at a higher redshift,
then the \ion{H}{2} bubbles are smaller and therefore more numerous at the
surface on bubble overlap which defines the end of reionization
\citep{wyithe04b}. Subsequently, this reduces the level of the Poisson
fluctuations, which is inversely proportional to the \ion{H}{2} region
number density.  The window function related to the finite size of a
typical \ion{H}{2} region, produces an exponential decay in the power
spectrum on smaller spatial scales. Thus the power spectra corresponding to
higher reionization redshifts, and therefore smaller characteristic sizes,
begin to exponentially decay at higher values of {\it l}.

In Fig. \ref{poisson} the amplitude of the polarization power spectra
increases with increasing reionization redshift. This conflicts with the
naive expectation that the polarization power spectra will be smaller at
high redshift since the Poisson fluctuations are smaller. The reason is
that the decrease in the characteristic \ion{H}{2} region size is rather
gradual and the visibility function is significantly increasing at high
redshifts. This is the main reason for the fact that the polarization power
spectra with low reionization redshifts have lower amplitudes. At yet
higher redshifts a turnover does occur and the decrease in characteristic
\ion{H}{2} region size overcomes the increase in optical depth.

The polarization power spectrum detection threshold due to instrument noise
is included on Fig. \ref{poisson} for the {\it Square Kilometer Array}
(SKA), which is the next generation low frequency radio interferometer.  We
adopt the power spectrum detection threshold model of \citet{zald04} with
updated parameters. We assume an instrument bandwidth of 
$\Delta \nu \sim 0.4 \mbox{ MHz}$ throughout the 
paper; increasing the bandwidth will decrease the instrument detection 
threshold as $\Delta \nu^{-1/2}$. 
In reality, the relevant power spectrum error bars for
a radio interferometer are a complcated function of both the exact
instrumental design and analysis method, so the detection thresholds in
this work should be viewed as a crude estimate (assuming that the
foregrounds can be properly removed).  Measurements of the proximity effect
around high redshift quasars implies that the end of reionization, i.e. the
surface of bubble overlap, is near $z = 6$ \citep{wyithe04a}. We find that
the polarization signal from even a redshift as low as $z_E = z_R = 6$ is
detectable with a 1-year integration time. A higher redshift of
reionization will only increase the amplitude of the polarization power
spectrum.

\begin{figure}
\plotone{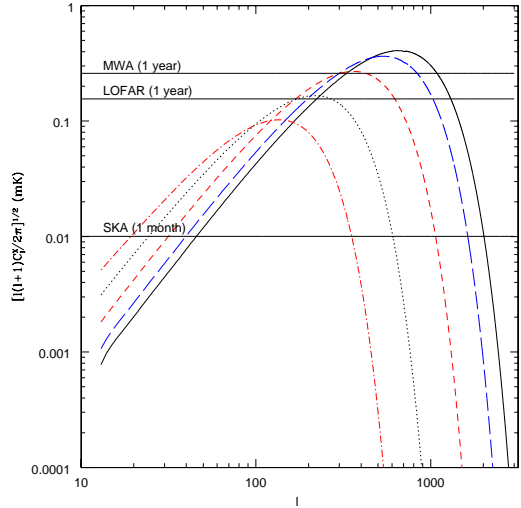}
\caption{The temperature-polarization cross-correlation spectra sourced by Poisson 
fluctuations of \ion{H}{2} regions for an emission and reionization redshift 
$z_E = z_R = 30$ (black, solid), $z_E = z_R = 25$ (blue, long dashed), $z_E = z_R = 17$ 
(red, dashed), $z_E = z_R = 10$ (black, dotted) and $z_E = z_R = 6$ (red, dot dashed). 
The theoretical detection threshold of various instruments : SKA - 1 month, LOFAR - 1 year 
and MWA - 1 year of integration time all with a bandwidth of $0.4 \mbox{ MHz}$.}
\label{cross}	
\end{figure}

In \S 2 we predicted that the temperature polarization cross-correlation
spectrum should be significantly larger than the polarization power
spectrum. Since the polarization power spectrum may only be detected by a
year long integration with a next generation instrument, we will compute
the cross-correlation spectrum and compare its amplitude to the noise
detection thresholds of upcoming instruments.  Figure \ref{cross} displays
the temperature-polarization cross-correlation spectra, sourced by
\ion{H}{2} region Poisson fluctuations, for different redshifts of emission
and reionization: $z_E = z_R = 30$ (black, solid), $z_E = z_R = 25$ (blue,
long dashed), $z_E = z_R = 17$ (red, dashed), $z_E = z_R = 10$ (black, dotted) and
$z_E = z_R = 6$ (red, dot dashed). Also shown are the instrument detection
thresholds for SKA - 1 month integration time, LOFAR - 1 year and MWA - 1 year.  
We can understand how the features
of these spectra differ from the polarization power spectra of
Fig. \ref{poisson} by comparing the formula for the temperature
polarization cross-correlation spectrum (Eq. \ref{xl}) with the formula for
the polarization power spectrum (Eq. \ref{HII_Cl}).

There are two obvious differences between the spectra. The temperature
polarization cross-correlation spectra have larger amplitudes and shallower
slopes on large scales. Clearly the polarization transfer function
(Eq. \ref{transfer}) has a smaller amplitude than the temperature transfer
function (Eq. \ref{tran_t}) because of the low optical depth, as well as,
the effects of the quadrupole free-streaming and conversion to E-type
polarization. The steep large scale slope observed in the polarization
power spectra (Figs. \ref{cl_hz}-\ref{poisson}) results from the quadrupole
nature of the production of polarization through Thomson scattering
\citep{hu97}. Each polarization transfer function roughly scales as ${\it
l}^2$ on large scales, so the polarization power spectrum should naturally
be steeper than the temperature polarization cross-correlation on these
scales. Once again, on small scales we observe the same exponential decay
caused by the \ion{H}{2} region window function. We find that even if
reionization ended at $z_R = 6$, this signal should be easily detectable by
SKA and very close to the detection threshold for LOFAR and MWA.  If new
observations of high redshift quasars imply that the surface of bubble
overlap occured at a higher redshift, then the detectability of the
temperature polarization cross correlation signal increases for LOFAR and
MWA. The result is very encouraging because it is based on simple
analytical considerations which are robust to variations in astrophysical
parameters. These uncertain parameters are all lumped into the value of
$z_R$ in our formulation.

\section{Discussion}

Figure \ref{cross} implies that the future {\it Square Kilometer Array}
(SKA) will have sufficient instrument sensitivity to detect the temperature
polarization cross-correlation power-spectrum of redshifted 21cm
fluctuations.  The forthcoming {\it Mileura Widefield Array} (MWA) and the
{\it Low Frequency Array} (LOFAR) might also have sensitivity to detect the
predicted signal. If new observations of high redshift quasars push this
redshift of bubble overlap to higher values then the possibility of a
detection increases.  The practical feasibility of such a detection will be
better known within a few years, as soon as the first polarized foreground
maps will be produced by LOFAR or MWA. The instrumental detection
thresholds used here are based on simplifications and should only be
interpreted as rough estimates.  However, the results shown in Figure
\ref{cross} portray the optimistic forecast that the polarization signal is
sufficiently large for it to be detectable.

We identified the dominant producer of polarization to be Thomson
scattering of an incident quadrupole moment of the 21cm radiation. The 21cm
quadrupole is sourced by the free-streaming of brightness fluctuations from
either correlated baryonic density fluctuations or Poisson fluctuations of
\ion{H}{2} regions. The amplitude of these Poisson fluctuations can be
robustly predicted based on the general considerations of cosmic variance
and light propagation delay at the end of reionization.
We have found that for a reionization redshift of $z_R = 6$, which is
suggested by the observations of the Gunn-Peterson effect, the temperature
polarization cross-correlation spectrum amplitude due to Poisson
fluctuations should be of order $\sim 0.1 \mbox{ mK}$ and the polarization
power spectrum amplitude $\sim 3 ~\mu \mbox{K}$. Strong cross-correlation
between the brightness temperature and polarization of the 21cm radiation
is generic for Thomson scattering. The cross-correlation will be weaker,
for example, if the polarization fluctuations are sourced by magnetic
fields whose distribution is spatially uncorrelated with the 21cm brightness
fluctuations.

Since Thomson scattering in a nearly uniform medium produces E-type
polarization and is achromatic, its unique fingerprints can be separated
from foregrounds associated with polarized synchrotron emission. Faraday
rotation of the signal as it propagates through the IGM will modify the above
conclusion, however the frequency dependence of the effect is precisely 
known and techniques developed to eliminate foregrounds may be adapted to 
reconstruct the original E-type polarization signal. 
It is highly unlikely that any foreground source would produce a polarized signal
that only contains E-type polarization and that correlates with the
cosmic brightness fluctuations. This may be used as a stringent test
against foreground contamination and instrument systematics. A similar test
has been adopted to check if the reconstructed signal from weak
gravitational lensing surveys is contaminated by instrumental systematics
and has proven to be extremely useful \citep{refregier03}. For upcoming experiments which may
not have the sensitivity to observe the polarized 21cm signal from the high
redshift IGM, the inferred signal should not correlate with any observed
polarization. This can be used to further eliminate the large foregrounds
that may contaminate the high redshift signal that is saught after.

\acknowledgements 

We would like to thank Matias Zaldarriaga, Oliver Zahn, Matt McQuinn, Chris Hirata,
Miguel Morales, Steve Furlanetto and Asantha Cooray for useful
conversations. This work was supported in part by NASA grants NAG 5-13292,
NNG05GH54G, and NSF grants AST-0071019, AST-0204514 (for A.L.).

\end{document}